# Can we track the geography of surnames based on bibliographic data?

Nicolas Robinson-Garcia[1], Ed Noyons[2] and Rodrigo Costas[2]

[1] *elrobin@ugr.es*
EC3Metrics spin-off and EC3 Research Group, Universidad de Granada, Granada (Spain)

[2] *noyons@cwts.leidenuniv.nl* [3] *rcostas@cwts.leidenuniv.nl*
Centre for Science and Technology Studies (CWTS), LeidenUniversity, Leiden (The Netherlands)

**Abstract**
In this paper we explore the possibility of using bibliographic databases for tracking the geographic origin of surnames. Surnames are used as a proxy to determine the ethnic, genetic or geographic origin of individuals in many fields such as Genetics or Demography; however they could also be used for bibliometric purposes such as the analysis of scientific migration flows. Here we present two relevant methodologies for determining the most probable country to which a surname could be assigned. The first methodology assigns surnames based on the most common country that can be assigned to a surname and the Kullback-Liebler divergence measure. The second method uses the Gini Index to evaluate the assignment of surnames to countries. We test both methodologies with control groups and conclude that, despite needing further analysis on its validity; these methodologies already show promising results.

**Conference Topic**
Data Accuracy and disambiguation

**Introduction**
Tracking the geographical origin of individuals has multiple applications and is of interest to many fields. For instance, in biomedical research it is used for racial and ethnic classification as this information is useful for identifying risk factors in epidemiological and clinical research (Burchard, et al., 2003). It is also of interest in the field of Demography to analyse migration movements (e.g. Chen & Cavalli-Sforza, 1983) or migratory influences in a given country (Hatton & Wheatley Price, 1999). In the field of bibliometrics, scientific migration flows between countries has been a subject of study as they are considered beneficial for the exchange of new ideas and scientific knowledge between countries (Moed & Halevi, 2014) as well as to analyse case studies to identify the spread of researchers of a given nationality around the world (Costas & Noyons, 2013).
Surnames have been used as a proxy of geographic, ethnic and even genetic origin for some time now. According to Kissin (2011) "the use of surnames in human population biology dates back to 1875, when George Darwin used frequency of occurrences of the same surname in married couples to study in-breeding". Geographic information related to surnames may also be of use in the field of bibliometrics, especially with regard to collaboration and mobility studies. So far only few papers have been found using surname data for bibliometric purposes. Kissin and colleagues (Kissin & Bradley, 2013; Kissin, 2011) have performed several studies focused on the analysis of Jewish surnames in the database MEDLINE. Also Freeman and Huan (2014) recently analysed the effect of diversity of authorship in the impact of scientific publications.
Until recently, these studies relied on manually curated lists of surnames related to ethnic groups, languages or countries. In the last few years, surname research has been developed and many methodologies have been proposed to discern statistical approaches to geographically classify surnames (a good review on the subject can be found in Cheshire, 2014). In this regard, two types of approaches can be found: 1) probability and Bayesian methods and 2) clustering techniques. For this, we can focus either on the concentration of surnames by areas or on tracking surnames to their original region (Cheshire, 2014).

So far the results reported are quite satisfactory (Mateos, 2007). While regional studies with large data sets offer relatively accurate results due to the skewness of the surnames distribution (Cheshire, 2014), there are still problems when applying these methodologies at a global level. Such limitations are due to migratory movements and data restrictions. For instance, the surname 'Lee' is considered in many studies as British. However, it is most common in the United States and at the same time in Asia. Also data availability may be an issue as most of it comes from census data and demography studies which usually come from different sources and present differences between them.

In this paper we suggest the use of a single data source to develop a methodology to track the geography of surnames worldwide. We propose using the authors' affiliation data from a scientific bibliographic database. For this purpose we analyse two different useful methodologies: one based on the application of information theoretic measures, and a second one based on the use of inequality indexes.

This paper is structured as follows. First we describe the data collection and processing. Then we describe each of the two methodologies proposed for assigning countries to names: one based on the Kullback-Leibler divergence (Kullback & Leibler, 1951) and a second one using the Gini Index, usually used in the field of Economics. In order to test the validity of each methodology, we compared our results with those from a list of surnames based on language origin for 11 different languages. Finally we conclude discussing the limitations of our methodologies, further developments and the potential use of this type of studies for the field of bibliometrics.

**Data collection and processing**

The goal of this paper is to develop a methodology to assign surnames to countries based on the bibliographic data offered by authors from a scientific database. For this we used the in-house CWTS version of the Web of Science database (not including the Conference Proceedings Citation Index or the Book Citation Index). This database covers all publications and authors for the 1980-2013 time period. The next step needed was to identify authors and relate them with their country of origin. Such approach assumes certain limitations:

- *Reliance on a single data source*. This means that errors or misrepresentations by countries derived from the Web of Science database will reflect on the quality of the result findings reported. Also, the surname information is restricted to the time period employed in the analysis, meaning that migration flows which have taken place before 1980 are not considered. This means that the origin of the surname is tracked according to a fixed image.
- *Limitations in the data*. We are working with a bibliographic database, implying that scholarly related patterns (e.g. migrations of scholars, mobility programs, issues related on how scholars use their name in publications, etc.) as well as database-coverage related problems (e.g. orientation of the database towards Anglo-Saxon countries, the lack of coverage of surnames that have never published, etc.) can play a role. Also, possible mistakes from the database (e.g., wrong linkage of authors to addresses, typos, transcription problems, lack of information, etc.) should be taken into account when interpreting the results.

In Figure 1 we offer an overview of the methodology followed. For all the surnames in all the publications covered in the Web of Science we detected all the 'trusted' linkages between authors and countries. By a trusted linkage we mean a surname-country relationship that is unambiguously registered in a publication[1] based on linkages between authors and countries according to bibliographic data. This implies that only in those cases where there is strong evidence that an author is linked to a country, the link is created and the combination (surname-country) is taken into consideration for the statistical analysis. These trusted linkages were created based on the following author-country combinations:

- Authors and countries from the *reprint address* field in the Web of Science are directly linked to their affiliation (Costas & Iribarren-Maestro, 2007).
- *Registered combinations of author and affiliations* recorded in the Web of Science, as from 2008 onwards WoS registers the linkage between authors and countries as they appear in the publications.
- *First authors* are assigned to the *first address* in the publication. As Calero and colleagues (2006) show the linkage of the first author with the first address of the publication is quite reliable.
- *One country publications*. For all publications with only one address or only national collaboration all their authors can be assigned to this country.

As a result, a matrix distribution of surnames by countries was created. Based on this matrix, two approaches were considered to assign surnames to countries. The first one consisted on assigning surnames to the countries with the highest frequency (in terms of publications containing the surname-country trusted linkage) which complied certain levels of assurance. This level of assurance was obtained by means of the Kullback-Liebler divergence or information gain measure. The second approach was to assign surnames according to their relative concentration by countries. This was done by using the Gini Index. In the next two subsections we detail each of the two methods proposed and the results obtained for each of them.

**Figure 1. Overview of the methodology followed for assigning countries to surnames**

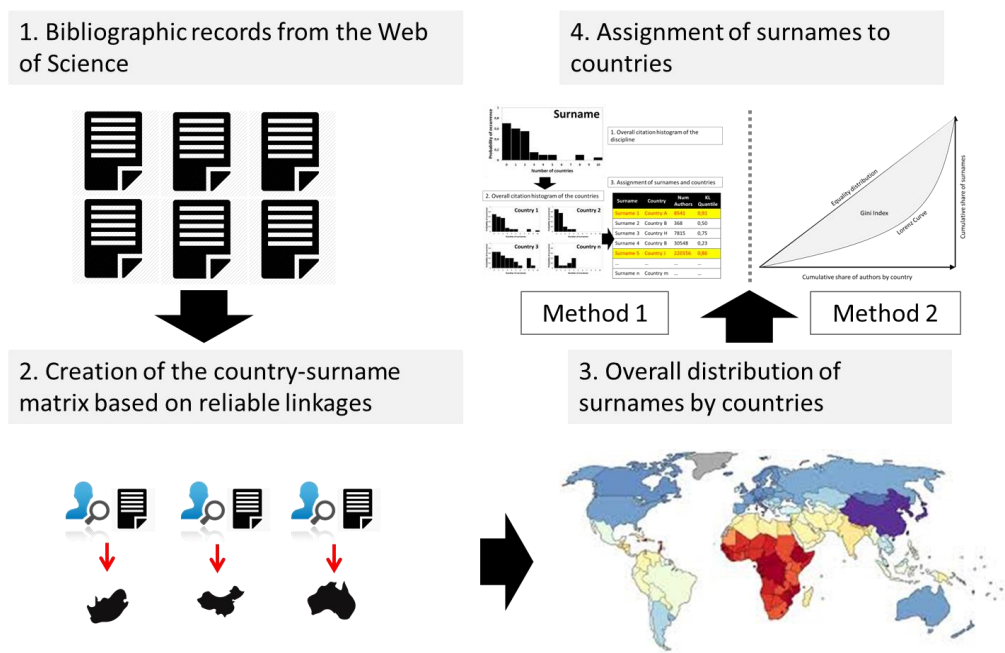

---

[1] For many publications in the Web of Science, not all the authors are directly linked to their affiliations in the paper, therefore sometimes it is very difficult to establish to which affiliation (and country) belongs every author.

**Method 1: Kullback-Leibler divergence and distribution by country**

When identifying the geographic origin of a surname one plausible approach is to consider that a surname will belong to the country in which the largest number of occurrences. However, this assumption entails two problems that have to be solved. Firstly, while using raw data will benefit countries with a large presence in the database (e.g. Western and Anglo-Saxon countries), relative indicators will benefit smaller countries, preventing from a balance between countries. Secondly, some surnames may show similar numbers in various countries. In order to overcome such limitations, we need a reasonable method to characterize the belonging of surnames to each country; and secondly, we have to be able to measure what is the amount of relative information between such characterizations. Here we propose the use of the information gain or Kullback-Leibler divergence measure (Kullback-Leibler, 1951). This measure allows us to select the country that contributes with more information to a given surname. It compares two distributions: a true probability distribution $p(x)$ and an arbitrary probability distribution $q(x)$, and indicates the difference between the probability of $X$ if $q(x)$ is followed, and the probability of $X$ if $p(x)$ is followed. Although it is sometimes used as a distance metric, information gain is not a true metric since it is not symmetric and does not satisfy the triangle inequality (making it a semi-quasimetric) (García et al., 2013).

In this paper, the true probability distribution $p(x)$ is represented by the authors' distribution of a given surname in the country with the highest number of such surname, while the arbitrary probability distribution $q(x)$ is represented by the frequency distribution of the surname in the rest of the countries. The objective is, on the one hand, to characterize the information gain between two probability distributions with a minimal number of properties which are natural and thus desirable. Second, it aims to determine the form of all error functions satisfying these properties which we have stated to be desirable for predicting surname-country dissimilarity. This analysis allows identifying similar and dissimilar distributions from a given one, but it does not explain the reasons for such dissimilarity. Such an approach has been previously used in the field of bibliometrics for very different purposes. For instance, Waltman and van Eck (2013) use it to identify national journals from international journals. García and colleagues (2013) use the Kullblack-Leibler divergence measure to determine similar academic institutions (García, et al., 2013). Finally, Torres-Salinas and colleagues (2013) apply it to characterize the field-specialization of publishers based on the citation patterns of book chapters (Torres-Salinas et al., 2013). In Figure 2 we summarize the main steps followed for assigning countries to surnames.

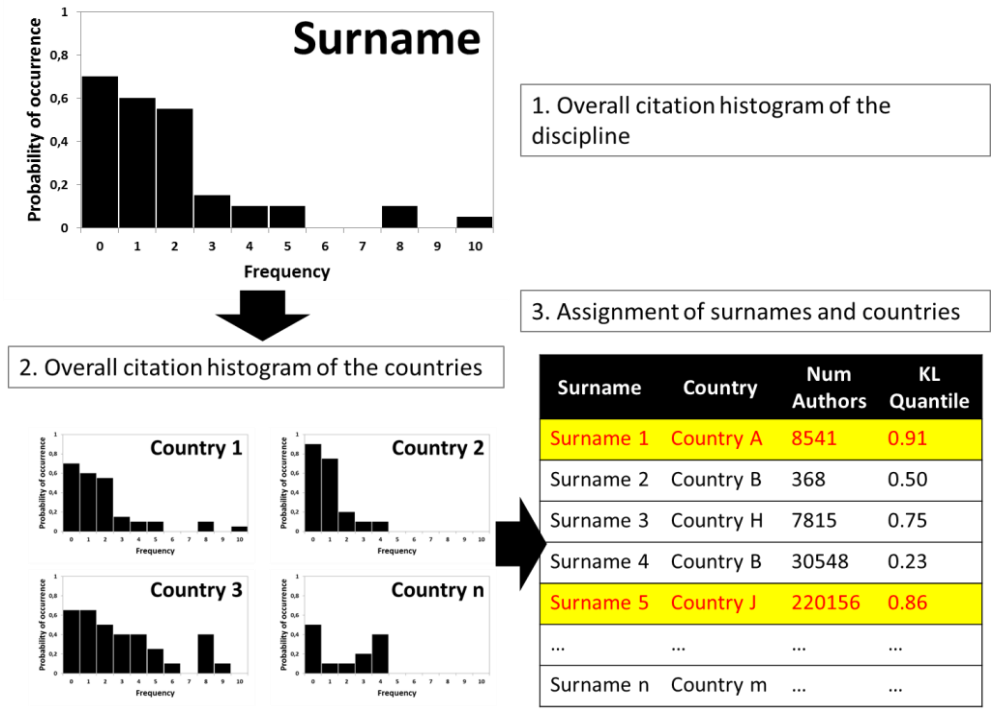

**Figure 2. Overview of Method 1 employing the Kullback-Leibler divergence measure**

If we predict the similarity between the given surname and the country based on their information gain, then we can set a minimum value of information gain that should be reached in order to ensure that the assignment made is correct, thus relating the surname with the country that leads to the most alike assignment to the frequency distribution. In this case we have established a minimum value up to the percentile $0.8^2$ of the overall distribution of surnames and main country by the Kullback-Leibler divergence measure in order to determine a good assurance in the surname-country association.

**Table 1. Distribution of top 36 countries with the highest number of surnames according to Method 1. Kullback-Leibler Divergence**

| Country | Surnames | Country | Surnames | Country | Surnames |
|---:|---:|---:|---:|---:|---:|
| FRANCE | 138349 | MEXICO | 38367 | FINLAND | 15160 |
| GERMANY | 112445 | BRAZIL | 37198 | UKRAINE | 14582 |
| RUSSIA | 111716 | GREECE | 34917 | CZECH REPUBLIC | 14427 |
| SPAIN | 83529 | IRAN | 34235 | NORWAY | 12892 |
| USA | 76219 | THAILAND | 32426 | DENMARK | 12861 |
| ITALY | 69637 | TURKEY | 27671 | ARGENTINA | 11714 |
| ENGLAND | 63885 | SWEDEN | 26134 | HUNGARY | 10541 |
| JAPAN | 56345 | ISRAEL | 24482 | PEOPLES R CHINA | 10472 |
| CANADA | 49775 | AUSTRALIA | 24259 | ROMANIA | 9976 |
| NETHERLANDS | 41306 | BELGIUM | 22203 | SOUTH AFRICA | 9504 |
| INDIA | 41198 | SWITZERLAND | 21402 | NIGERIA | 9313 |
| POLAND | 40446 | AUSTRIA | 18048 | EGYPT | 8682 |

---

[2] In other words, we consider that up to 80% of the surname-country linkages based on the highest KL divergence measures are informative, and we disregard 20% of the combinations in which the surname and the country cannot be considered as a reliable linkage (as the surname could also reasonably belong to another country, based on the overall distribution of the surname across countries).

*Results*

A total of 1,568,052 surnames were assigned to 119 different countries. Table 1 shows the distribution by surnames of the 36 countries with the higher number of surnames assigned. As observed, the largest number of surnames is assigned to the France (8.8%), followed by Germany (8.0%), Russia (7.1%) and Spain (4.9%).

As observed, some countries with the same language appear in this list, such as England and United States for English language or Spain and Mexico for Spanish language. Also some manual normalization of countries was required due to changes in the name of countries (i.e., USSR and Russia or Germany and Federal Republic of Germany).

**Method 2: Gini inequality index and concentration by country**

Another plausible approach to assigning countries to surnames is to consider the right country as the one where a given surname is more concentrated. For this, we suggest the use of inequality indexes such as the Gini Index. This indicator has already been used in the field of bibliometrics. For example, Torres-Salinas and colleagues (2014) employ it to determine the level of specialization of academic publishers indexed in the Book Citation Index. It is a measure of statistical dispersion. It is defined based on the Lorenz Curve, which plots the proportion of population (y axis, surnames in our case) that is cumulatively concentrated by the bottom $x$% of the population. In Figure 3 we represent its interpretation. The equality distribution is represented by a 45 degrees line. The Gini Index is defined as the ratio of the area that lies between the line of equality and the Lorenz Curve. Its value ranges between 0 and 1, 0 meaning total equality (or dispersion) and 1, total inequality (or concentration). The hypothesis we pose is that a surname can be assigned with certain levels of reliability to the country which shows a higher concentration of such surname, hence relativizing the presence of a given country in the database.

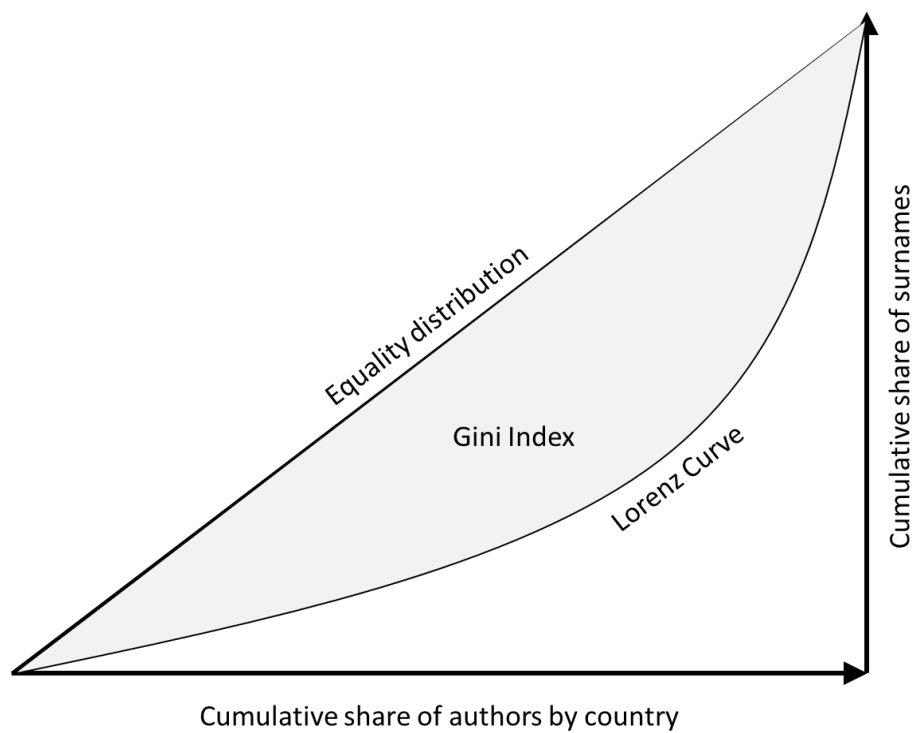

**Figure 3. Interpretation of the Gini Index**

In Table 2 we show the distribution of surnames by countries for the top 36 countries with the highest number of surnames. A total of 1,885,782 surnames were matched to a list of 343 countries. The country with the largest number of surnames assigned is the United States, representing 16.5% of the total share, and followed by France (6.25%) and Germany (5.9%). In general terms we observe that this methodology distributes surnames among a larger number of countries, showing a less skewed distribution.

**Table 2. Distribution of top 36 countries with the highest number of surnames according to Method 2. Gini Index**

| Country | Surnames | Country | Surnames | Country | Surnames |
|---|---|---|---|---|---|
| USA | 310739 | NETHERLANDS | 40528 | UKRAINE | 17580 |
| FRANCE | 117938 | BRAZIL | 38386 | ARGENTINA | 16275 |
| GERMANY | 111375 | GREECE | 38034 | FINLAND | 16060 |
| RUSSIA | 94369 | IRAN | 37162 | CZECH REPUBLIC | 15166 |
| SPAIN | 77387 | THAILAND | 35090 | NORWAY | 15074 |
| ITALY | 65699 | TURKEY | 28473 | DENMARK | 14347 |
| JAPAN | 52399 | ISRAEL | 28360 | HUNGARY | 12291 |
| ENGLAND | 47521 | SWEDEN | 26051 | ROMANIA | 11767 |
| CANADA | 46146 | SWITZERLAND | 25029 | SOUTH AFRICA | 11018 |
| POLAND | 44087 | BELGIUM | 23863 | NIGERIA | 10619 |
| INDIA | 42897 | AUSTRALIA | 23396 | CHINA | 9531 |
| MEXICO | 41066 | AUSTRIA | 21609 | EGYPT | 9158 |

**Validation**

In order to validate the results of each method and determine their performance, we tried to compare them with a 'valid' list of surnames by countries. However, identifying such a list entails certain limitations. First, there is no 'perfect' and unique linkage between countries and surnames. Secondly, these linkages are not usually done for countries but rather for languages, cultures, ethnicities, etc. We decided to use a list of surnames by language provided from Wikipedia[3] and select a sample of languages.

**Table 3. Control table of correspondences between countries and languages**

| Normalized country | Languages | Countries |
|---|---|---|
| Denmark | Danish | Denmark; Greenland |
| England | Celtic; Anglo-Cornish; English; Scottish; Irish | Antigua & Barbuda; Australia; Bahamas; Barbados; Belize; Bermuda; Canada, England, Ghana; Gibraltar; Grenade; Guyana; Ireland; Jamaica; Liberia; Malawi; Mauritius; Micronesia; N Wales; Namibia, New Zealand, Nigeria; Scotland; Sierra Leone; Solomon Islands; South Africa, St. Kitts & Nevis; St. Lucia; St. Vincent; Trinidad & Tobago; USA; Wales; Zambia |
| Finland | Finnish | Finland |
| France | Breton; French | Benin; Burkina Faso; Congo; Côte Ivoire; Polynesia; France; French Guayana; Gabon; Guadeloupe; Guinea; Haiti; Ivory Coast; Mali; Martinique; Monaco; New Caledonia; Niger; |

---
[3] http://en.wikipedia.org/wiki/Category:Surnames_by_language

| Normalized country | Languages | Countries |
|---|---|---|
| | | Reunion; Senegal; Togo; Upper Volta |
| Germany | German | Austria; Germany; Liechtenstein |
| Greece | Greek | Greece |
| Iceland | Icelandic | Iceland |
| Italy | Italian | Italy; San Marino; Vatican |
| Japan | Japanese | Japan |
| Netherlands | Afrikaans; Dutch | Holland; Netherlands; Surinam |
| Portugal | Portuguese | Angola; Brazil; Cape Verde; Guinea Bissau; Mozambique; Portugal |
| Spain | Basque; Catalan; Galician; | Andorra; Argentina; Bolivia; Chile; Colombia; Costa Rica; Cuba; Dominican Republic; Ecuador; El Salvador; Guatemala; Honduras; Mexico; Nicaragua; Panama; Paraguay; Peru; Spain; Uruguay; Venezuela |

We chose 20 different languages grouped in what we called 12 'normalized' countries, that is, the most representative countries of these 20 languages. Then we crossed our sample table with the surnames obtained from Web of Science and identified the countries to which each of the two methods proposed assigned these surnames. The list of countries was then processed in order to identify the 20 languages selected. We assigned to each retrieved country one of the selected language if one of the following premises was given (Table 3):

1. It was the official language of the country. For instance, French is the official language of countries such as Gabon, Haiti or Martinique.

2. It is not the main language but it is only spoken in a given area. For instance, Galician, Basque and Catalan surnames were assigned to Spain, or Breton to France.

3. There is more than one official language (which is also used in other countries). This is the most important limitation noted from our validation method, as it excludes countries such as Switzerland, Belgium or Luxembourg (which have several languages spoken in more than one country). The only exception noted is Canada, which has been attributed to English language, acknowledging the important limitation towards French language.

Our validation list from Wikipedia contains a total of 8,239 surnames. After crossing this list with our list of surnames retrieved from Method 1, a total of 7,625 surnames were matched. In Table 4 we include the distribution of surnames by normalized countries according to our control list (Table 3), the coverage of 'valid' assignments made, that is, those surnames which could be assigned with certain levels of assurance according to their information gain; and the share of correct assignments.

**Table 4. Distribution of surnames by countries of the control sample for 12 normalized countries according to their language, valid assignments and correct assignments according to the two methods proposed**

| | | METHOD 1* | | | METHOD 2** | |
|---|---|---|---|---|---|---|
| Countries | Surnames | % coverage | % correct | Surnames | % coverage | % correct |
| DENMARK | 123 | 91.06% | 68.75% | 123 | 100% | 60.16% |
| ENGLAND | 932 | 28.76% | 80.97% | 929 | 100% | 58.56% |
| FINLAND | 225 | 99.11% | 94.62% | 224 | 100% | 91.96% |
| FRANCE | 562 | 88.08% | 68.28% | 560 | 100% | 50.54% |
| GERMANY | 2186 | 52.24% | 69.00% | 2170 | 100% | 43.78% |

|            | **METHOD 1*** | | | **METHOD 2*** | | |
| ---: | ---: | ---: | ---: | ---: | ---: | ---: |
| Countries | Surnames | % coverage | % correct | Surnames | % coverage | % correct |
| GREECE | 170 | 84.12% | 78.32% | 168 | 100% | 78.57% |
| ICELAND | 29 | 100.00% | 65.52% | 28 | 100% | 100.00% |
| ITALY | 972 | 87.65% | 86.97% | 968 | 100% | 64.77% |
| JAPAN | 1349 | 98.74% | 98.95% | 1347 | 100% | 91.39% |
| NETHERLANDS | 471 | 88.11% | 60.96% | 468 | 100% | 41.67% |
| PORTUGAL | 137 | 98.54% | 92.59% | 136 | 100% | 91.91% |
| SPAIN | 469 | 93.18% | 48.74% | 464 | 100% | 54.74% |
| **Total** | **7625** | **73.22%** | **79.03%** | **7585** | **100%** | **61.29%** |

\* **Method 1:** Kullback-Leibler divergence; ** **Method 2:** Gini Index

As observed, in general terms the coverage of 'reliable' assignments made was of 73.2% of the sample list. However, significant differences can be found by country. While in the case of Iceland all surnames were assigned with certain levels of assurance (>80 quartile of the Kullback-Leibler divergence distribution), in the case of England only 28.8% of the surnames were considered valid. Also the coverage figures are quite low for Germany (52.2%). From these surnames covered, around 80% of them were assigned to the correct country. The highest figures of correct assignments are observed for Japan (98.9%, also with a coverage of 98.5%), while the lowest figures go to Spanish surnames (48.7% of correct assignments with a coverage of 93.2%). In the case of England, although the coverage is low, 80.1% of the assignments were correct. In the case of Germany the share is lower (69%).

Regarding the methodology based on the Gini Index, a total of 7585 surnames were retrieved after crossing the list of surnames obtained with the control list. As observed, the coverage of 'reliable' assignments with this methodology is much higher (100%), however, many differences are observed on the share of correct assignments. In general terms this methodology performs not as well as the first one, with 61.2% of all assignment correct. However, in some cases its share of correct assignments is higher. This is the case of Iceland where the 29 surnames of the control list were correctly assigned. Also the share of correct assignment for Spain increases (54.7%) but still shows low values.

**Discussion and conclusions**

In this paper we propose the identification of the geographic origin of surnames for bibliometric purposes. For this, we propose the use of scientific databases in order to work with data worldwide. In this way we overcome a major restriction of this type of studies regarding data availability (Cheshire, 2014). We propose two methodologies to assign countries to surnames. The first method is based on the number of surnames found in a given country when its Kullback-Leibler divergence measure is below the $80^{th}$ percentile of all the combinations with the lowest Kullback-Leiber values. The second methodology is based on the concentration of a given surname in a country, using the Gini Index to calculate such concentration.

In this regard, a preliminary validation has been done comparing the coverage and correct assignments made with a sample list of 20 languages grouped into 12 'normalized country'. The results reported are promising, especially for the first methodology. In fact, this has already been applied successfully elsewhere (Costas & Noyons, 2013). But the second methodology ensures a 100% coverage of all surnames. However, much research is still needed and further refinements in both methodologies. First, we believe that thresholds of minimum publications of a surname by country should be established in order to improve the methodology based on the Gini Index. Regarding the Kullback-Leibler divergence methodology, we considered reliable assignments those which were below the 80[th] percentile, however, different thresholds should be also tested. Finally, we will consider other validation lists as some questionable assignments were found in this control list (e.g., Pinto is assigned to Italian language, but it could also be assigned to Spanish or even Portuguese) which may blur the evaluation of the actual performance of each method. These methods should also be compared with those developed elsewhere.

The use of surnames to track demographic movements or analyse diversity in collaboration shows interesting opportunities for implementing these methodologies in bibliometric analyses. One example of such application is the recent work of Freeman and Huan (2014). However, frequently little attention to the methodology employed for assigning countries, languages or ethnicities to surnames is paid, something that may represent a challenge to results based on these data. Thus, understanding better the limitations and possibilities of these data is critical for a proper use. Although further research is still needed, we believe that applying methodologies such as the ones suggested here using bibliographic databases will lead to more reliable results.